\documentclass{elsarticle}
\usepackage{amsmath}
\usepackage{graphicx} 
\usepackage{bbold}
\usepackage{algpseudocode}
\usepackage{algorithm}
\usepackage[margin=1in]{geometry}
\usepackage{url,color}
\usepackage[colorlinks,linkcolor=blue,citecolor=blue,urlcolor=blue]{hyperref}

\begin{document}

\title{A partitioned shift-without-invert algorithm to improve parallel eigensolution  
efficiency in real-space electronic transport} 

\author{Baruch Feldman \fnref{1}}  
\author{Yunkai Zhou\fnref{2}}

\fntext[1]{Department of Materials and Interfaces, Weizmann Institute of Science, Rehovoth 76100, Israel. 
{\tt Email:~baruch.feldman@weizmann.ac.il}.
Supported in part by the European Research Council and the Israel Science Foundation.}
\fntext[2]{Department of Mathematics, Southern Methodist University, Dallas, Texas, USA, 75093. {\tt Email:~yzhou@smu.edu}.
Supported in part by NSF grants DMS-1228271 and DMS-1522587.}

\begin{abstract}
We present an eigenspectrum partitioning scheme without inversion for the recently described real-space electronic transport code, TRANSEC.  The primary advantage of TRANSEC is its highly parallel algorithm, 
which enables studying conductance in large systems. 
The present scheme adds a new source of parallelization, significantly enhancing TRANSEC's parallel scalability, especially for systems with many electrons.  In principle, partitioning could enable super-linear parallel speedup, as we demonstrate in calculations within TRANSEC.  In practical cases, we report better than five-fold improvement in CPU time and similar improvements in wall time, compared to previously-published large calculations.
Importantly, the suggested scheme is relatively simple to implement. It can be useful for general large Hermitian 
or weakly non-Hermitian eigenvalue problems, whenever relatively accurate inversion 
via direct or iterative linear solvers is impractical.
\end{abstract}

\maketitle

\section{Introduction}

Recently a real-space approach has been developed for Green's function-based ab-initio electronic conductance calculations, called TRANSEC \cite{TRANSEC}.  
TRANSEC inherits a number of intrinsic advantages associated with real-space electronic-structure calculations, including favorable parallelizability and no requirement of an explicit basis-set \cite{TRANSEC, cts:94, parsec-review}.  

Within this approach the bottleneck in computing the electronic transmission function $T(E)$ at energy $E$ is the partial diagonalization of a complex-symmetric matrix according to the equation
\begin{equation}
\left( H_{KS} - i \Gamma \right) \: U_k \; = \; \epsilon_k \: U_k \: , \; \epsilon_k \: \in \: \mathbb{C} \ .
\label{eq:complex-energy-evp}
\end{equation}
As described in detail previously \cite{TRANSEC}, we perform this step as an abbreviated, cheaper intermediate to the inversion  
\[
G(E) \; \equiv \; \left\{ E \mathbb{1} - (H_{KS} - i \Gamma) \right\}^{-1} \; , 
\] 
which is 
needed to compute transmission: 
\[
T\left( E \right) = \mbox{Tr}\left\{ G\left( E \right) \: \Gamma_R \: G^\dagger\left( E \right) \: \Gamma_L \right\} .  
\]  
In Eq.~(\ref{eq:complex-energy-evp}), $H_{KS}$ is the Hermitian Kohn-Sham (KS) Hamiltonian obtained from the 
PARSEC Density Functional Theory (DFT) code \cite{cts:94, parsec-review} after self-consistency is reached;
the $i \Gamma \equiv i (\Gamma_L + \Gamma_R)$ is a sum of imaginary, diagonal absorbing potentials with Gaussian form at the two ends of the simulation cell, where $i^2 = -1$;  
and the $\epsilon_k$ and $U_k$ are a pair of eigenvalue and eigenvector, respectively, of $H_{KS} - i \Gamma$ \cite{TRANSEC}.  
To calculate conductance in an implicit real-space basis, we use a simulation cell having a finite volume $V$, so the dimension of the matrix $H_{KS}$ is given by $N \approx V / h^3$.  Here $h$ is the grid spacing of the real-space lattice and $N$    
is typically $10^4$ to $10^6$ or greater. 
To correct for the finite volume $V$, the imaginary absorbing potentials, $i \Gamma$, are tuned to absorb outgoing electrons and prevent reflections at the boundaries of the simulation cell \cite{TRANSEC}.  Since $H_{KS}$ is real-symmetric, the presence of $i \Gamma$ results in a complex-symmetric eigenproblem.   
This partial diagonalization 
is the most computation-intensive part of TRANSEC, and can take many hundreds or thousands of core-hours of computation on standard supercomputers \cite{TRANSEC}.  

TRANSEC is parallelized mainly by partitioning the real-space grid over computing cores during matrix-vector application\footnote{Note that symmetry and $k$-point sampling are not implemented because of the non-periodic geometry of conductance calculations, so parallelization over these dimensions is unavailable.}.  This monolithic source of parallelization results in less-than-optimal parallel scaling when, for instance, the number $N_e$ of electrons under simulation increases in tandem with, or even faster than, the supercell volume $V$ \cite{TRANSEC}.  

In this paper, we develop a further dimension of parallelization by partitioning over the 
eigenspectrum of Eq.~(\ref{eq:complex-energy-evp}).  This scheme can significantly improve the iterative solution of the eigenproblem (\ref{eq:complex-energy-evp}), leading to markedly better scalability in TRANSEC.  As we will demonstrate in Section~\ref{sec:applications}, super-linear parallel speedup, or 
net savings in computing time, are possible.  
The present scheme is 
conceptually and operationally simpler than shift-invert \cite{Meerbergen:1997:IRA,appinv-eig00,sip07,silan-14} and similar algorithms, yet still effective.  
It can accelerate the solution of large eigenvalue problems, 
both Hermitian and non-Hermitian, when matrix inversion via
matrix decompositions such as sparse LU or via iterative linear solvers is impractical.   
We emphasize that (\ref{eq:complex-energy-evp}) is weakly non-Hermitian\footnote{
By ``weakly,'' we mean both that the anti-Hermitian part $- i \Gamma$ is restricted to the diagonal, and that it is typically small, as discussed around Eq.~(\ref{eq:evalue-PT}).   
}, as 
the eigenvectors of (\ref{eq:complex-energy-evp}) are bi-orthogonal
with respect to a standard inner product \cite{TRANSEC, Santra}; in contrast, for
Hermitian eigenvalue problems, the eigenvectors are orthogonal. 
While iterative methods have been studied \cite{CSYM, complex-Lanczos} for solution of linear equations with complex-symmetric matrices, few have been proposed for the corresponding eigenvalue problems, such as \eqref{eq:complex-energy-evp}. 
Robust production code for the solution of  sparse large-scale 
complex-symmetric eigenvalue problems such as Eq.~(\ref{eq:complex-energy-evp}), 
especially when many eigenpairs are needed, is sorely missing.
The algorithm we propose in this paper can be used to solve such  
complex-symmetric eigenvalue problems. 

This paper is organized as follows.  The remainder of this section introduces  
the no-inverse approach and outlines some of its advantages.  
In Section~\ref{sec:theory} we describe in greater detail 
the shift-without-invert method,  
including comparisions to other   
partitioned approaches, and some heuristics towards an automated partition method.  
In Section~\ref{sec:applications} we present several large-scale calculations 
illustrating the potential for improved parallelization and reduced computational time of 
the proposed method,
including two major applications originally reported in \cite{TRANSEC}.
Our current calculations show 
better than five-fold net savings in computational time compared to our previous method.

The method proposed here partitions the eigenvalue problem (\ref{eq:complex-energy-evp}) into parallel sub-problems, then rigidly shifts the operator
\begin{equation}
H_{KS} - i \Gamma
\label{eq:operator}
\end{equation} into 
each different partition of the eigenspectrum. 
In contrast to more commonly used shift-invert approaches, 
such as \cite{Meerbergen:1997:IRA,appinv-eig00,sip07,silan-14},  
we avoid matrix inversions.  Thus our algorithm involves 
no linear equation solving via $LU$ decomposition or iterative linear equation solvers.

Existing spectrum partition algorithms (e.g. \cite{sip07,silan-14})  
partition the large eigenproblem into parallel sub-problems, but must 
still face the complexity associated with shift-inverse.
For our problem, the matrix (\ref{eq:operator}) is very large,
and furthermore, the matrix $H_{KS}$ in TRANSEC is never explicitly computed or stored 
\cite{TRANSEC, cts:94}, even in a standard sparse format; it is instead 
represented as a matrix-vector product formula.    
These facts make inversion \textit{via} $ILU$ factorization infeasible 
(even if it may be possible to store
the sparse $H_{KS}$ explicitly, the cost of factorization can still be prohibitive due to the matrix size). 
Sparse inversion using an approximate inverse (AINV) preconditioner 
\cite{SaadBook,benzi-precond02} may be another option, 
but both the AINV and the ILU preconditioners are not straightforward 
to apply; 
moreover, even with the 
increased coding complexity and the expected increase in memory cost, 
the overall CPU time is not necessarily reduced.

TRANSEC works with the Hamiltonian $H_{KS}$ from PARSEC, which is defined on a 
real-space grid \cite{cts:94}.  
The computational cost of the iterative diagonalization of Eq.~(\ref{eq:complex-energy-evp}) 
using a Krylov-type method such as ARPACK 
scales like $O(N n_r^2)$, where 
$n_r$ is the total number of eigenpairs found \cite{arpack,govl:96}.
Because of the sparsity of the Hamiltonian from a real-space method, multiplying the Hamiltonian by a trial vector contributes  
only linearly in the Hamiltonian dimension $N$ to the cost of \eqref{eq:complex-energy-evp}. 
Yet this matrix-vector application is the critical source of parallelization in real-space methods \cite{cts:94}. 
By contrast, the quadratic scaling in $n_r$ is associated with orthogonalizing the growing 
subspace of eigenvectors, as well as with subspace rotations necessary to carry out 
the Rayleigh-Ritz refinement process \cite{pchefsi}.  
Whereas the Hamiltonian dimension depends on both the system volume $V$ (number of atoms) 
and the grid spacing $h$ as $N \approx V/h^3$, the number of eigenpairs $n_r$ needed to converge the $T(E)$ calculation typically scales only with the number $N_e$ of electrons in the system.  
Because the system volume, number of atoms, and number of electrons typically scale together, 
the cost of our original method \cite{TRANSEC} 
grows cubically with system size $V$, whereas the parallelizability improves only linearly with $V$.  
Worse still, parallel scalability 
deteriorates when a large $n_r$ is needed for a given $N$, for example in systems with a high average electron density per unit volume, $N_e/V$.  

To overcome these problems, we present here 
a spectrum partitioning method 
that enables the solution of Eq.~(\ref{eq:complex-energy-evp}) with better parallel scalability.
To avoid the problems associated with inverse of large matrices mentioned above, 
we approach this problem by shifting without inversion.  
The no-inverse approach allows us to parallelize our transport method over energy as well as 
over the real-space grid, significantly improving the overall parallel performance, 
as will be shown in  Section \ref{sec:applications}.

\section{The shift-without-invert partition algorithm \label{sec:theory}}  

Note that we need to compute the $n_r$  eigenvalues with smallest-real-part\footnote{
The reason is that to first order, conductance is a property of a small range of eigenpairs around the Fermi energy $E_F \equiv \epsilon_{N_F}$, the $N_F$th-lowest eigenvalue in the spectrum of $H_{KS}$ \cite{TRANSEC}.  Typically in TRANSEC we choose $n_r$ a few times $N_F$, and $N_F \sim N_e/2 \ll N$, where $N_e$ is the number of electrons in the simulation, and $N \approx V/h^3$ is 
the dimension of $H_{KS}$. 
Hence in practice the required eigenpairs are identical to 
the $n_r$ eigenpairs with smallest-real-part. 
(In this paper any ``ordering'' of the eigenvalues always refers to ordering by the real parts.)   
\label{note:EF}
}  
and their associated eigenvectors of
Eq.~(\ref{eq:complex-energy-evp}).
To partition this computation, one main feature of the present scheme is to shift 
the matrix (\ref{eq:operator}) by some strategically placed real rigid shifts $E_{S,j}$  
and transform Eq.~(\ref{eq:complex-energy-evp})
into a sequence of $p$ sub-problems: 
\begin{equation}
  \left( H_{KS} - i \Gamma - E_{S,j} \right) \: U_k \; = \; (\epsilon_k - E_{S,j}) \: U_k \: , \; 1 \leq j \leq p \: , 
  \label{eq:evp-shift}
\end{equation}
which can be solved parallelly. 
For each shift $E_{S,j}$, we solve for  $n_{r,j}$ number of eigenvalues closest to the shift, where each $n_{r,j}$ is only a fraction of $n_r$. 
After combining a sequence of such partitions $j$ with different values of the shifts $E_{S,j}$ 
and solving for $n_{r,j}<n_r$ eigenpairs on each partition, we obtain the desired total $n_r$ eigenpairs.
To compute all $n_r$  
eigenpairs, it is clearly necessary that 
\begin{equation}
\sum^p_j \: n_{r,j} \: \geq \: n_r \:. 
\label{eq:sum-nr}
\end{equation}
The number $n_{r,j}$ of eigenvalues computed in each sub-problem can in theory be made much smaller than the total $n_r$, hence we mitigate the 
$O(N n_r^2)$ complexity into a sequence of $p$ parallel sub-problems, each with only $O(N n_{r,j}^2)$ complexity. 

By breaking (\ref{eq:complex-energy-evp}) into a sequence of mostly independent sub-problems, the present scheme greatly enhances the parallelization of our real-space method, as we will demonstrate.  Moreover, partitioning allows ``continuation'' runs to expand a domain of previously-solved eigenpairs from $n_r$ to a total $n_r' > n_r$, 
by computing just the $n_r' - n_r$ previously unsolved eigenpairs as a new partition.

A single partition's complexity is quadratic in $n_{r,j}$, 
thus it is theoretically possible to solve for a total of $n_r$ eigenpairs 
in a time that scales linearly in $n_r$, i.e. $O(n_r)$, when we partition the spectrum 
into as many finer parts as necessary.  In particular, this in theory can lead to far  
better than the ``ideally'' linear parallel speedup of non-partitioned methods.     
In practice, though, our approach faces challenges owing to the 
increased computational demand associated with computing interior eigenpairs.  
Our scheme must also cope with uncertainty in choosing the shifts $E_{S,j}$, 
since the eigenvalues and their density distribution are unknown {\it a priori}.

The applications we present in Section \ref{sec:applications} illustrate in practice both the potential, and the possible setbacks, of the present partitioning scheme.  These results  
show the possibility of super-linear parallel speedup, or equivalently, saving CPU time \textit{via} partitioning, even compared to un-parallelized calculations.  In particular, Sections \ref{sec:au111} and \ref{sec:BDT} present two major applications containing Au(111) nanowires from Ref.~\cite{TRANSEC}, each of which we compute here in far less CPU time than in Ref.~\cite{TRANSEC}, on the same hardware.  
The third major application from Ref.~\cite{TRANSEC}, a C$_{60}$ molecule between Au(111) leads, 
is not presented here. In this case, one interior partition out of four had difficulty converging using the same options passed to
PARPACK.

The main reason is that Lanczos-type methods (including PARPACK without inversion) are better suited to converging exterior eigenvalues.  For
eigenvalues located far interior to the spectrum, Lanczos-type
methods may suffer from slow convergence, which can worsen when the interior eigenvalues are clustered.  With our partitioned method, we call PARPACK with the `SM' (smallest
magnitude) option, iterating over a much smaller dimension subspace compared to a non-partitioned method.  Thus our approach is most effective when eigenvalues are not highly clustered\footnote{
Of course, other factors like the assignment of grid points to cores, numerical roundoff, and convergence tolerance also influence convergence in practice.  
}.  For the unconverged partition in the C$_{60}$ test application, the requested number of eigenpairs per Ry is over 2200. In comparison, the average requested eigenpairs per Ry for the Au nanowire examples is less than 1340, while for the C chain it is less than 250.
The more eigenpairs requested per Ry, the more clustered some interior
eigenvalues can become.  This helps explain why we encountered
convergence difficulty for one partition in C$_{60}$ yet had excellent results for the
other applications.  
For problems with highly clustered interior eigenvalues, a pure no-invert
approach as developed here may not be optimal.  In this case we should resort to 
applying shift-with-invert, that is, apply 
the more complicated `inverting' techniques only when the no-invert approach encounters difficulty converging some partitions.  
This combination of no-invert and invert will be of future development.  
We emphasize that, for many problems in quantum transport where the interior eigenvalues are not highly 
clustered, our no-invert approach provides a greatly simplified alternative to shift-invert.

We implement the shift-without-inverse scheme to study quantum transport 
with the TRANSEC \cite{TRANSEC} code. 
Here, 
the anti-Hermitian part $- i \Gamma$ of (\ref{eq:operator}) is a relatively small perturbation to $H_{KS}$, 
so the imaginary parts $\Im\{\epsilon_k\}$ of the eigenvalues are bounded, and much 
smaller than the magnitudes of the real parts,
\[ 
|\Im\{\epsilon_k\}| \; \ll \; |\Re\{\epsilon_k\}| .
\]  
Specifically,
the complex eigenvalues in our TRANSEC applications are given by
\begin{equation}
\epsilon_k = \langle U_k | H_{KS}-i \Gamma | U_k \rangle \; \approx \; \epsilon_k^{KS} - i \langle U_k^{KS} | \Gamma | U_k^{KS} \rangle. 
\label{eq:evalue-PT}
\end{equation} 
Here the $U_k^{KS}$ are unperturbed KS eigenvectors (i.e., those of $H_{KS}$ alone), 
\[
\epsilon_k^{KS} \equiv \langle U_k^{KS} | H_{KS} | U_k^{KS} \rangle \; \in  \; \Re
\]
are the corresponding unperturbed KS eigenvalues, and the approximate equality can be justified by first-order perturbation theory.  
Therefore  
\[
0 \: \leq \: -\Im\{\epsilon_k\} \: \leq \: \max\{\Gamma\}, 
\]  
that is, 
the whole spectrum lies near the real axis, 
and TRANSEC's eigenproblem Eq.~(\ref{eq:complex-energy-evp}) is only weakly non-Hermitian. 
Consequently, we only find it necessary to choose appropriate shifts 
$E_{S,j}$ on the real axis. This choice of real shifts  simplifies the partitioning of the spectrum.

It is worth mentioning that our scheme should also be applicable to more strongly non-Hermitian eigenproblems.  
In such cases one needs to choose complex shifts to compute eigenvalues with larger imaginary parts, and a two-dimensional partitioning $\{ E_{S,j,k} \: , \: n_{r,j,k} \}$, 
where the $j$ and $k$ indexes represent real and imaginary parts of the shifts,    
should be necessary to cover the whole eigenspectrum.

\subsection{Rationale of a ``without-invert'' approach \label{sec:rationale}}

Our partitioned algorithm belongs to the divide-and-conquer methodology. As is well-known,
a divide-and-conquer method in theory is more suitable for parallel computing 
than non-partitioned counterparts.
The rationale behind our partitioned method for eigenvalue problems is also to exploit the
potential parallel scaling efficiency. More specifically, since a standard 
non-partitioned sparse iterative eigenalgorithm for computing $n_r$ eigenpairs of a dimension
$N$ matrix has complexity $O(Nn_r^2)$, if we partition the wanted spectrum into $p$ parts, 
then for each part the complexity reduces to 
$O(N \left(\frac{n_r}{p}\right)^2)$, so in theory the total
complexity reduces to  
\begin{equation}
p \cdot O\left(N \left(\frac{n_r}{p}\right)^2 \right) \: = \: O\left(N \; \frac{n_r^2}{p}\right) \: .  
\label{eq:theoretical-complexity}
\end{equation} 
We note these theoretical results reference only the number $p$ of partitions, not the number of cores, so (\ref{eq:theoretical-complexity}) 
could theoretically be attained even with $p$ serial runs on a single core.  But there is very little cost to parallelizing 
the shift-without-invert scheme because the $p$ partitions can be computed in embarassingly 
parallel (i.e., entirely independently of each other), as indeed was assumed to derive Eq.~(\ref{eq:theoretical-complexity}).  The only coordination or communication required among partitions is to combine the results after the eigensolution steps, and subsequently to fill in any missed eigenpairs at the interfaces among partitions, as described below.  
Therefore, we may also consider $p$ as the number of
available processors or CPU cores, which on a modern supercomputer can readily reach 
over a few thousand (or we could choose $p$ larger than the number of cores by running additional parts in series in order to benefit from the $\sim 1/p$ scaling).  So the ideal complexity could reach even  
\begin{equation}
O\left(N n_r\right) \: ,   
\label{eq:linear-complexity}
\end{equation} 
when $p =O(n_r)$. 

The superior theoretical scaling efficiency associated with partitioned methods 
is the driving force behind the eigenvalue partition algorithms, represented by
\cite{sip07},  \cite{slicing-12}, and \cite{silan-14}.

Both Refs.~\cite{sip07} and \cite{silan-14} utilize the shift-invert operation within the framework
of the Lanczos method; the earlier such decomposition idea traces back to \cite{lanczos-para87}.
The inverse operations require solving linear equations, which are usually
realized by calling either an iterative linear equation algorithm such as CG, or a sparse direct 
solver \cite{davis-book}, in particular the MUMPS package \cite{mumps}.  
However, it is well-known that solving 
the shifted linear equations involves significant efforts, both in computational cost\footnote{
In fact, we have in the first place chosen eigensolution as an efficiency-enhancing intermediate step toward our actual goal of inverting $E - (H_{KS} - i\Gamma)$, as discussed around Eq.~(\ref{eq:complex-energy-evp}) and further in Ref.~\cite{TRANSEC}.}  
and especially in
the algorithm development.
A sparse direct solver is the more stable and more straightforward 
choice when the dimension is not large; however,
for very large dimension linear equations, a direct solver becomes impractical and
an iterative solver often is needed. But any iterative linear equation solver
is not a black-box method that can be used straightforwardly.  Rather, there are several parameters related
to a given solver that must be tuned \cite{temp:lin,petsc-ug}, 
and such task of choosing parameters is further complicated 
by the fact that the shifted equations are often ill-conditioned. The so-called 
``preconditioned'' eigen-algorithms are usually more complicated to use than the preconditioned linear
solvers they employ to solve the shifted linear equations. 

Furthermore, the Lanczos method employed in \cite{lanczos-para87, sip07, silan-14}
may suffer from missing eigenvalues, especially when there are clustered eigenvalues and when
subspace-restart is used to save memory; therefore, quite complicated post-processing operations are
required to find any eigenvalues missed by a previous Lanczos run within a partition. 
To guarantee no eigenvalues are missed in a partition, accurate eigenvalue counts on 
an interval must be calculated in \cite{sip07, silan-14}. This is done by resorting to the 
Sylvester inertia theorem, which would require a sequence of Cholesky decompositions.
The $O(N^3)$ complexity associated with a Cholesky decomposition thus can significantly restrict 
the dimension of eigenvalue problems to which the partitioned methods in  \cite{sip07, silan-14}
can be applied.

To avoid these difficulties associated with inversion, 
the spectrum slicing method in \cite{slicing-12} opts to apply Chebyshev-Jackson polynomial filters.
However, very high order of degree (such as $\ge 1000$) polynomials are 
used to 
achieve the desired filtering, resulting in significant computational cost for the
filtering. The method in
\cite{slicing-12} seeks to avoid expensive Cholesky decompositions for counting 
eigenvalues.  This saves computational cost, but increases the complexity of the
algorithmic design, since several post-processing procedures are needed to guarantee finding all eigenvalues on a partition slice.
The method in \cite{slicing-12} is applied to only a relatively small number of partitions;
extending it to many partitions could encounter difficulties, 
owing to the complexity of applying high order degree polynomial filters, and the 
requirement of finding all eigenvalues when eigenvalue counts are unknown.

As discussed earlier, the main advantage of a partitioned eigenvalue algorithm should be its
applicability to as many partitions as possible, therefore, 
we adopt a different approach here than in \cite{slicing-12} to avoiding inversion.
In addition, the partitioned eigen-methods cited above are all restricted to Hermitian 
eigenproblems, whereas our method is applicable to non-Hermitian eigenproblems, 
such as Eq.~\eqref{eq:complex-energy-evp} in TRANSEC.

Although the scheme we propose here could be used with any $O(N n_r^2)$ iterative eigensolver, 
our implementation makes use of the well-received eigenvalue package 
ARPACK \cite{arpack, website:arpack-ng}, which arguably remains the best public domain solver for large
non-Hermitian eigenvalue problems.  
ARPACK can solve a standard eigenvalue problem such as (\ref{eq:complex-energy-evp}) 
without performing inverse, 
simply by applying implicit 
polynomial filters \cite{sorens:92}. 
In TRANSEC we actually call PARPACK -- the parallel version of ARPACK using MPI.

As mentioned earlier, the $O(Nn_r^2)$ scaling complexity of PARPACK leads to  
inefficiency when $n_r$ is large. 
Our solution is to decompose the spectrum into smaller chunks and solve parallelly 
on each chunk for only a small number of eigenvalues.  
When computing a relatively small number of eigenvalues, PARPACK enjoys
excellent scalability due to fewer basis vectors needing re-orthogonalization  
(and thus fewer inner-products). This, coupled with the overall stability of ARPACK, 
makes PARPACK the best available choice for our partitioned subproblem \eqref{eq:evp-shift}. 
In \eqref{eq:evp-shift}
the number $n_{r,j}$ of requested eigenvalues is only a very small fraction of the dimension of 
the Hamiltonian matrix.  

However, we encounter an immediate difficulty with the partition approach: although PARPACK provides 
options to specify which eigenvalues to compute, such as SR/LR (smallest/largest real part) and SM/LM (smallest/largest magnitude),  
the SR/LR/LM are all for computing exterior eigenvalues, and cannot be used to compute 
the eigenvalues in, for example, an interior partition; and the remaining SM option is only for
computing eigenvalues closest to zero.

We overcome this difficulty by combining shifts and the SM option in PARPACK. That is, we 
strategically place shifts as in Eq.~\eqref{eq:evp-shift} in an estimated region of the spectrum, 
then request PARPACK to compute the eigenvalues closest to each of 
these shifts by using the SM option on the shifted operator. 
With this choice, we can converge both the exterior and the interior eigenvalues,
by placing shifts at suitable locations of the spectrum. 
A downside is that some interior eigenvalues may be very slow to converge 
without using inverse; if that happens, the overall scalability of our scheme would 
deteriorate.  
A possible remedy is to partially integrate inverse operations when such a situation is detected to 
happen. Partially utilizing inverse is viable 
because the number of eigenvalues to be computed around a shift is small. 
The mixed no-inverse plus shift-inverse approach is still expected to be 
less expensive than using shift-invert on the full wanted spectrum, and will be the subject of our future work.  
The current paper focuses on the shift-no-inverse approach.

\subsection{Computational structure of the shift-without-invert algorithm \label{sec:main-structure}}

Existing partitioned eigenvalue algorithms, including \cite{sip07, slicing-12, silan-14} and our
method, all face the challenge of how to partition the spectrum.
Since the spectrum is unknown at the start of computation, it is not straightforward to know how to 
partition it into smaller parts, and harder still to partition in such a way that each chunk
would have similar workload for ideal load balancing. 
This is one of the intrinsic difficulties of any partitioned approach; another difficulty
is the handling of partition boundaries (or interfaces) between adjacent partitions, 
including removing redundant eigenvalues and
recovering the wanted eigenvalues that may have been missed on all the partitions. 
These difficulties, in our opinion, may be the main reason why there exist 
rather few partitioned eigen-algorithms, 
even though such an approach can theoretically reach excellent linear scaling 
complexity.  
Progress toward an automated algorithm to cope with these difficulties would be a 
meaningful step forward for approaching the ideal theoretical scaling efficiency promised
by a partitioned method.

In the next sections 
we suggest simple techniques to address the two intrinsic difficulties mentioned above. 
Note that the approach as it is sketched here is neither optimized, 
nor robust enough to run entirely without human input. Thus our approach is not yet a fully automated algorithm.
But the present heuristics are straightforward to implement and 
may serve as an initial step toward a fully automated partitioned eigen-algorithm.
Note also that the results in Section~\ref{sec:applications} below are special in that we obtained them with foreknowledge of the eigenspectrum from our previous work \cite{TRANSEC}, and therefore did not need to resort to these approximate methods to estimate the $n_{r,j}$ and $E_{S, j}$.

\subsubsection{Choosing shifts and partitions \label{sec:choosing-heuristic}}

We choose in advance the total number $p$ of partitions, based on how many cores are available for parallelization, and based on the general consideration that performance improves as $p$ increases (as seen for actual calculations in Section \ref{sec:applications}) until the overhead of choosing the partitions begins to dominate.   
Each partition is associated with a shift $E_{S, j}$ together with a number  
$n_{r,j}$ of eigenpairs to solve; in the $j$-th partition, we compute the $n_{r,j}$ eigenpairs closest to the shift
$E_{S, j}$.

To estimate initial locations for the partitions, we call PARPACK to compute just two
exterior eigenvalues -- the one with the largest real part and the one with the smallest. 
This is not expensive since PARPACK is generally very
efficient for converging a small number of exterior eigenvalues.  Our goal in TRANSEC 
is to compute the lower end of the spectrum (see footnote \ref{note:EF}), so once we estimate these two eigenvalues 
(denote their real parts as $r_{\min}$ and $r_{\max}$), 
we can estimate the interval to be partitioned 
as a certain proportion of the full 
eigenspectrum located at the lower end, for example:  
\[ 
\left[ r_{\min}, \; r_{\min} + \frac{n_r}{N} \cdot \left( r_{\max} - r_{\min} \right) \right] .
\] 
However, this estimate assumes uniform density of the eigenspectrum, so it could lead to too many or too few eigenpairs within the partitioned range. 
To improve, 
we can estimate the density of eigenpairs   
by solving for a few eigenpairs at several 
sample points throughout the spectral region of interest.  

We next discuss the challenge of load balancing.  Because PARPACK is based on the restarted Arnoldi / Lanczos method, 
interior eigenvalues are more expensive to converge than exterior ones when we call it without inversion.  
Clearly, using equal $n_{r,j} \equiv n_r / p$ on each partition $j$ would result in poor load balancing.  
Instead, it is strongly advisable to reduce $n_{r,j}$ as the partitions move  
into the interior of the spectrum.
As a simple heuristic, we use the formulas at lines \ref{nrj1}-\ref{nrj2} 
in the pseudocode shown in the next section to do this. 
After choosing $n_{r,j}$,  
we must estimate $E_{S,j}$ accordingly, as for example in line \ref{ESj}.   
Here {\tt{round()}} and  {\tt{ceil()}} are the standard rounding and 
ceiling functions, respectively.
The formulas, derived by approximate fitting to actual timing data, provide cheap estimates of the optimal $n_{r,j}$ for given 
$j$ and $p$ that result in far improved load balancing.  
Still, these formulas are only heuristics, so we expect further refinements 
such as sampling the local eigenvalue density would 
improve load-balancing significantly.

We now address the two issues mentioned above relating to boundaries between partitions:  
first, some eigenvalues may be computed twice 
in adjacent partitions, resulting in redundancy.   
Second, some wanted eigenvalues may be outside the range of all partitions, leading to holes in
the computed spectrum. 
In practice, these two issues compete; tolerating slight 
redundancy may be preferable so as to minimize the risk of holes.  
Holes are the more severe problem because filling them necessitates further PARPACK calls.  

To address the redundancy problem, we first identify eigenpairs having the same eigenvalue within a small numerical tolerance, and then
perform a Gram-Schmidt process to bi-orthogonalize their computed eigenvectors.
As a result of the Gram-Schmidt process, any eigenvector linearly dependent on the previous eigenvectors is removed, in this case we also remove its associated redundant eigenvalue.  
The search and subsequent Gram-Schmidt process can 
run through all eigenpairs, or only over adjacent pairs of partitions.\footnote{
In theory, a brute-force search for degenerate eigenvalues across all partitions could cost quadratically in $n_r$, just as does the orthogonalization step in PARPACK's Arnoldi algorithm.  
But one must carry out this search and the Gram-Schmidt step only 
as post-processing steps {\em after} each solution of Eq.~\eqref{eq:evp-shift}, rather than iteratively within Eq.~\eqref{eq:evp-shift}.  
So in practice, removing redundancy consumes only a small fraction of CPU time.  
Moreover, one can restrict this search to adjacent partitions, 
since redundancy normally does not extend beyond the nearest partition.  }

The problem of missing eigenvalues is harder to address.
Since we avoid any Cholesky decompositions, we cannot apply the Sylvester inertia theorem 
to get eigenvalue counts on a partition. Aside from the high cost of  
Cholesky decompositions, our eigenvalue problems are non-Hermitian, so the 
Sylvester inertia theorem does not apply.  In addition, TRANSEC and PARSEC avoid explicitly computing or storing $H_{KS}$ or other $N \times N$ matrices, so
a matrix decomposition such as Cholesky is inapplicable. 
Some techniques have been recently proposed \cite{spectral-dos,int-count}
to estimate the number of eigenvalues 
in a given interval without computing them. 
But these provide only approximate counts, and are again intended for Hermitian eigenproblems,
thus cannot apply to \eqref{eq:evp-shift}.  
Next, we instead propose simple heuristics to identify and fill holes.  

\subsubsection{Pseudocode of the partitioned shift-without-invert algorithm}\label{sec:algorithm}

Here we present first heuristics towards more automated hole-filling approaches.
The structure of our algorithm is presented as Algorithm \ref{algo:partition}, which 
we have designed with the quantum transport application of TRANSEC in mind. 
Adjustment may be called for to optimize the algorithm for other applications.  

Some useful inputs to the algorithm include the total dimension $N$ of the Hamiltonian and the total eigenspectrum range $r_{min}$ and $r_{max}$, as mentioned above.  In quantum transport applications, one can also use the lowest Kohn-Sham eigenvalues $\epsilon_k^{KS}$ (those of $H_{KS}$ only, without $i \Gamma$) to estimate the eigenvalue spacing and 
the lowest eigenvalue of \eqref{eq:operator}.  These are necessary prerequisites to quantum transport, obtained when solving the KS equations of DFT, and are valid approximations to $\epsilon_k$, as discussed around Eq.~\eqref{eq:evalue-PT}.  Still, the algorithm would benefit from non-automatic human insight to gauge the validity of such an approximation, or further corrections such as higher-order perturbation theory to improve on it.

Following Algorithm \ref{algo:partition}, one first can use the $n_{r,j}$ from lines \ref{nrj1}-\ref{nrj2} 
to estimate optimal shifts $E_{S,j}$, as in line \ref{ESj}.  Ideally, these shifts would be chosen to minimize redundancy while still avoiding 
holes, both between partitions, and at the edges of the overall range of interest.  
Or one can improve on line \ref{ESj} by estimating the eigenspectrum distribution, as described above.  
Any such knowledge of the spectral distribution could prove important to the quality of the initial shift choices.

Next, one computes the solution to Eqs.~\eqref{eq:evp-shift} using the chosen $n_{r,j}$ and $E_{S,j}$.  
The Gram-Schmidt process shown in Algorithm \ref{algo:combine} should then be performed to remove redundancies.  As mentioned above, in practice a few redundancies are desirable to minimize holes.  
In fact, we find a reliable heuristic to detect holes is the absence of even a single redundancy between adjacent partitions.  
If the Gram-Schmidt process detects holes, we next apply additional iterations to ``fill in'' the holes,
 by inserting new partitions between existing adjacent ones that lack redundancy, as in lines \ref{fill-hole-1}-\ref{fill-hole-2} of Algorithm \ref{algo:partition}.  
This approach is reminiscent of an adaptive grid algorithm.

Hole-filling carries 
computational costs and overheads, but these must be weighed against 
the more expensive probing that would be 
needed to get an accurate count of the number of eigenvalues in advance.

\begin{algorithm}  
\caption{Partitioned shift-without-invert eigen-algorithm, with automated hole-filling heuristic} \label{algo:partition}
\begin{algorithmic}[1]   
	\State $p :=$ Total number of partitions, $\{ E_{min}, E_{max} \} := $ extremal eigenvalues  
	\For{$j = 1 \to p$}  
			\State Estimate number of target eigenpairs $n_{r,j}$ for partition $j$: 
					\If{$j \leq p/2$ }  \label{nrj1}
						\State $n_{r,j} = {\tt{round}}\left( \frac{2}{1.5 + \frac{j-1}{p}} \cdot \frac{n_{r}}{p} \right)$  
					\Else
						\State $n_{r,j} = {\tt{ceil}}( \frac{2 n_{r}}{p} ) - n_{r,(p-j+1)} $
					\EndIf  \label{nrj2}
			\State Estimate energy shift $E_{S,j}$ for partition $j$: $E_{S,j} = E_{S,j-1} + \; \left( \frac{E_{max} - E_{min}}{N} \right) \left( \frac{n_{r,j} + n_{r,j-1}}{2} \right) $  \label{ESj}
			\State Solve eigenvalue problem (\ref{eq:evp-shift}) 
                         with the updated values $n_{r,j}$, $E_{S,j}$
	\EndFor  

		\\
  \State Set a conditional flag ``holes'' on each interface between partitions to true 
	\Repeat  {(while any ``holes'' flag is true)}
		\State Combine partitions and check for holes (Algorithm \ref{algo:combine}) 
		\\
		\ForAll{interfaces $j$ on which the flag has value ``holes''} \label{fill-hole-1} 
			\State Create a new partition $j'$ between partitions $j-1$, $j$; \; $p = p + 1$; 
\State Choose $E_{S,j'} = \frac{1}{2} \left[ \max_i \Re\{\lambda^{(j-1)}_{i}\} + \min_i \Re\{\lambda^{(j)}_i\} \right]$ , 
where $i$ indexes all eigenvalues $\lambda^{(k)}_i$ on partition $k$; 
\State Choose $n_{r,j'} = \frac{3}{2} \left[ \frac{ \min_i \Re\{\lambda^{(j)}_i\} - \max_i \Re\{\lambda^{(j-1)}_{i}\} }{(E_{max} - E_{min})/N} \right] $ , 

			\State Solve (\ref{eq:evp-shift}) using $n_{r,j'}$, $E_{S,j'}$.  

		\EndFor \label{fill-hole-2}
	\Until{No ``holes'' remain}.
\end{algorithmic}
\end{algorithm}

\begin{algorithm}
\caption{Combine partitions and remove redundancies / test for holes} \label{algo:combine}
\begin{algorithmic}[1] 
	\ForAll{Computed eigenpairs $\{\lambda_k, v_k\}$, across all partitions}
		\State Check whether eigenvalue $\lambda_m = \lambda_k$ for any $m < k$ : 
			\ForAll{Computed eigenpairs $m < k$}
				\If{$\lambda_m = \lambda_k$}  
\State Modified Gram-Schmidt orthogonalization: 
~$v_k = v_k - {\tt Proj}_{v_m} v_k$
				\EndIf
			\EndFor   
		\State Compute norm $a_k \equiv |v_k|$ after Gram-Schmidt process; 
		\If{$a_k < $ tolerance} 
			\State Remove redundant eigenpair $\{ \lambda_k, v_k \}$ ; 
			\State Set flag holes$_{k-1} = $ false ;  
		\Else
			\State Normalize: $v_k = v_k / a_k$ ;
		\EndIf

	\EndFor
\end{algorithmic}
\end{algorithm}

\section{Application of the partitioned shift-without-invert algorithm \label{sec:applications}}

In this Section, we present benchmark timing results of the shift-without-invert scheme compared to the standard single-partition TRANSEC method of Ref.~\cite{TRANSEC} for several large $T(E)$ calculations.

We define parallel speedup $\eta$ according to the usual convention, except with two generalizations.  First, because the partitions $j$ run independently, we report speedup either based on total CPU time, or on the longest elapsed wall-time:
\begin{equation}
\eta_{_{CPU}} \equiv \frac{T_{CPU,0}}{T_{CPU}} \; , \; 
\eta_{wall} \equiv \: \frac{T_{wall,0}}{\max_j\{T_{wall,j}\} \: \cdot N_{cores}} \; .  
\label{eq:speedup}
\end{equation}
Here $T_{CPU}$ is the total CPU time for the parallel run, $T_{CPU,0}$ is the total CPU time for the reference serial run, $T_{wall,j}$ is the elapsed wall time of the $j$th partition, $T_{wall,0} \approx T_{CPU,0}$ the elapsed wall time of the reference serial run, and $N_{cores}$ the total number of cores in the parallel job (cumulative over all partitions).  
Note the difference between $\eta_{_{CPU}}$ and $\eta_{wall}$  
is due to imperfect load-balancing among the partitions, i.e. $T_{wall,k} \neq T_{wall,j}$. 
Second, because some calculations are too large to run practically on a single core, 
we sometimes replace the single-\emph{core} reference times $T_{CPU,0}$ and $T_{wall,0}$ in Eq.~(\ref{eq:speedup}) with single-\emph{partition} reference times $T_{CPU,1}$ and $N_1 \cdot T_{wall,1}$, respectively, that are still parallelized over grid points to $N_1$ cores.  

For ideal parallelization, the speedup factors defined above approach $\eta_{_{CPU}} = \eta_{wall} =$ 100\%.  In this Section, we will report cases where $\eta > $ 100\% because multiple partitions can actually \emph{reduce} CPU time compared to a single partition, as discussed 
around Eqs.~(\ref{eq:theoretical-complexity}) and (\ref{eq:linear-complexity}).  

Our shift-without-invert scheme requires a relatively small additional run to combine the results of the separate partitions (Algorithm \ref{algo:combine}).  We neglect this in most timings reported in this section, but including it would not change the qualitative picture we present.  
A more important caveat is the necessity to choose $E_{S,j}$ and $n_{r,j}$ appropriately, something made harder when one lacks foreknowledge of the distribution of eigenvalues.  
As mentioned in Section~\ref{sec:main-structure}, the calculations in this Section portray the potential of a somewhat idealized shift-without-invert scheme because we do possess such foreknowledge from Ref.~\cite{TRANSEC}. 
Therefore, hole-filling was avoided, and  
the timings we present here are simply cumulative times for the successful partitioned runs.  But we note the calculations \emph{do} reflect 
realistic difficulties such as load-balancing and redundancy. 

In general, we sought to equal or exceed the total eigenvalue counts $n_r$ of Ref.~\cite{TRANSEC}, choosing $n_{r,j}$ in accordance with Eq.~(\ref{eq:sum-nr}), and using lines \ref{nrj1}-\ref{nrj2} of Algorithm \ref{algo:partition} as a starting point for load-balancing\footnote{When doubling the number of partitions, this typically means that each partition $j$ is split into two smaller ones.}.  
We divided 
the single-partition eigenspectra obtained in Ref.~\cite{TRANSEC} into consecutive intervals $j$ containing $n_{r,j}$ eigenvalues, and chose $E_{S,j}$ as the midpoint of each interval.  
We then typically increased the actual eigenpair requests $n_{r,j}$ 
by $\sim$5\% or 10\%, and in 
some cases\footnote{
The C chain and the 2- and 4-partition BDT calculations were successfully partitioned on the first attempt.} 
further adjusted the partitions, in order to eliminate holes.  
We combined the partitions according to Algorithm \ref{algo:combine}, and proceeded to compute transmission $T(E)$ as in Ref.~\cite{TRANSEC}.  
Because $T(E)$ is sometimes sensitive to the number of eigenpairs used \cite{TRANSEC}, we typically might discard eigenpairs in the combined spectra that were in excess of the number found in the corresponding single-partition runs.  
In addition to the $T(E)$ comparisons shown below, we also usually compared the final 
eigenvalue spectra to the single-partition results as another rigorous consistency check.  

\subsection{C monatomic chain \label{sec:C-chain}}

\begin{figure}[!ht]
\begin{center}
\includegraphics[scale=0.35]{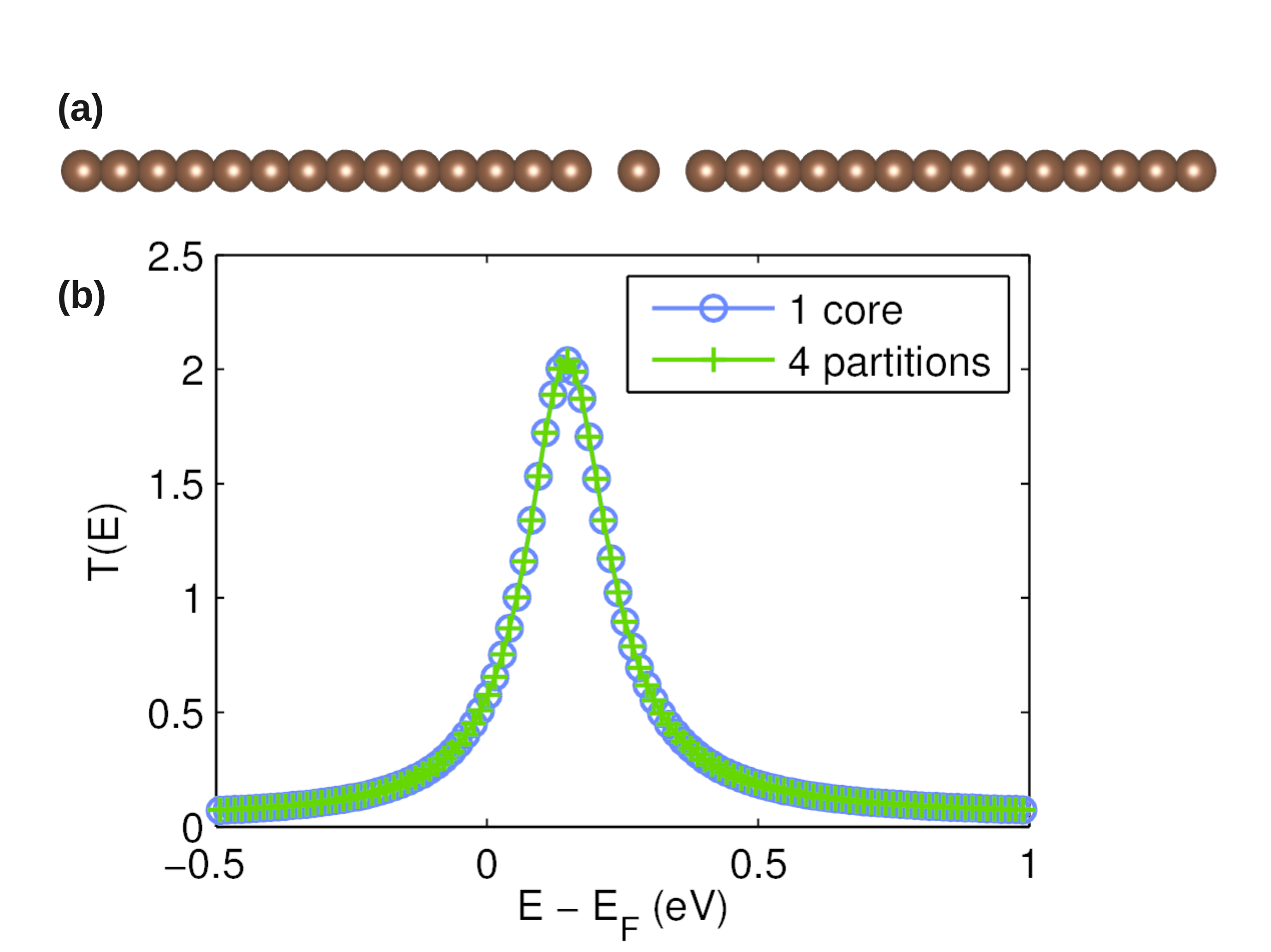} 
\caption{(a) Structure of the C monatomic chain system considered here and in Ref.~\cite{TRANSEC}.  In this work we use 10\% of total eigenpairs rather than 5\%.  (b) TRANSEC calculated results for transmission, $T(E)$, for the C monatomic chain system shown in (a).  Results obtained using the partitioned shift-without-invert scheme agree with the single-partition results to better than two decimal places.  
\label{fig:C-chain} }
\end{center} 
\end{figure}

We applied Algorithm \ref{algo:partition} in TRANSEC to compute the transmission $T(E)$ in an identical C monatomic chain structure 
presented in Ref.~\cite{TRANSEC}. 
The geometry, shown in Fig.~\ref{fig:C-chain}(a), consisted of 14 C atoms per electrode, with atomic spacing of 2.6 $a_0$, and a gap of 4.7 $a_0$ between the electrodes and central atom, where $a_0$ is the Bohr radius.  The calculation made use of norm-conserving Troullier-Martins pseudopotentials with \textit{s}/\textit{p} cutoff radii of 1.46/1.46 $a_0$ for C.  As in Ref.~\cite{TRANSEC}, we used Gaussian imaginary absorbing potentials $\Gamma$ centered on the first and last atoms in the chains, of strength 265 mRy and standard deviation 10.4 $a_0$.  

Following Ref.~\cite{TRANSEC}, we used a converged grid spacing $h=$ 0.6 $a_0$, resulting in 
$N=$ 22,100 grid points.  In order to investigate a case with large $n_r/N$, we requested the lowest $n_r/N=$ 10\% of eigenpairs instead of 5\% as in Ref.~\cite{TRANSEC} for a total of $n_r=$ 2,210 eigenpairs. 
The eigenvalues range from $E_F \: -$ 1.036 Ry to $E_F \: +$ 9.229 Ry,
where   $E_F$ is the Fermi level.
On a single core with all $n_r$ eigenvalues in a single partition, the calculation took about $T_0 =$ 14.5 hours.  Parallelized to 4 cores via the standard TRANSEC procedure (i.e., a single partition, parallelized over $N$ only), it took $T_{wall,1} =$ 10 wall-clock hours, 
or about $T_{CPU,1} =$ 40 core-hours (parallel speedup of just $\eta =$ 36\%).  

With the shift-without-invert algorithm with 4 partitions on 1 core each (total of 4 cores), 
the calculation took a total of $T_{CPU} =$ 9 core-hours, giving parallel speedup $\eta_{_{CPU}} =$ 160\% compared to the single-core run. 
The result was obtained in wall-clock time $\max_j\{T_{wall,j}\} =$ 3 hours and including Algorithm \ref{algo:combine}, a full $T(E)$ result was obtained within 3.5 hours of starting the calculation (parallel speedup of $\eta_{wall} =$ 290\% compared to the non-partitioned parallel run with same 4 cores, or $\eta_{wall}>$100\% compared to the single core).  
The partition parameters were $n_{r,j} = (760,\: 630,\: 459,\: 424)$ and $E_{S,j} - E_F = (1.726,\: 5.678,\: 7.496,\: 8.678)$ Ry.  
The $T(E)$ results agree with the results obtained with the non-partitioned PARPACK package 
in Ref.~\cite{TRANSEC}, 
as shown in Fig.~\ref{fig:C-chain}(b).

\subsection{Transmission in Au(111) nanowire electrodes} \label{sec:au111}

\begin{figure}[!h]
\begin{center}
\includegraphics[scale=0.43]{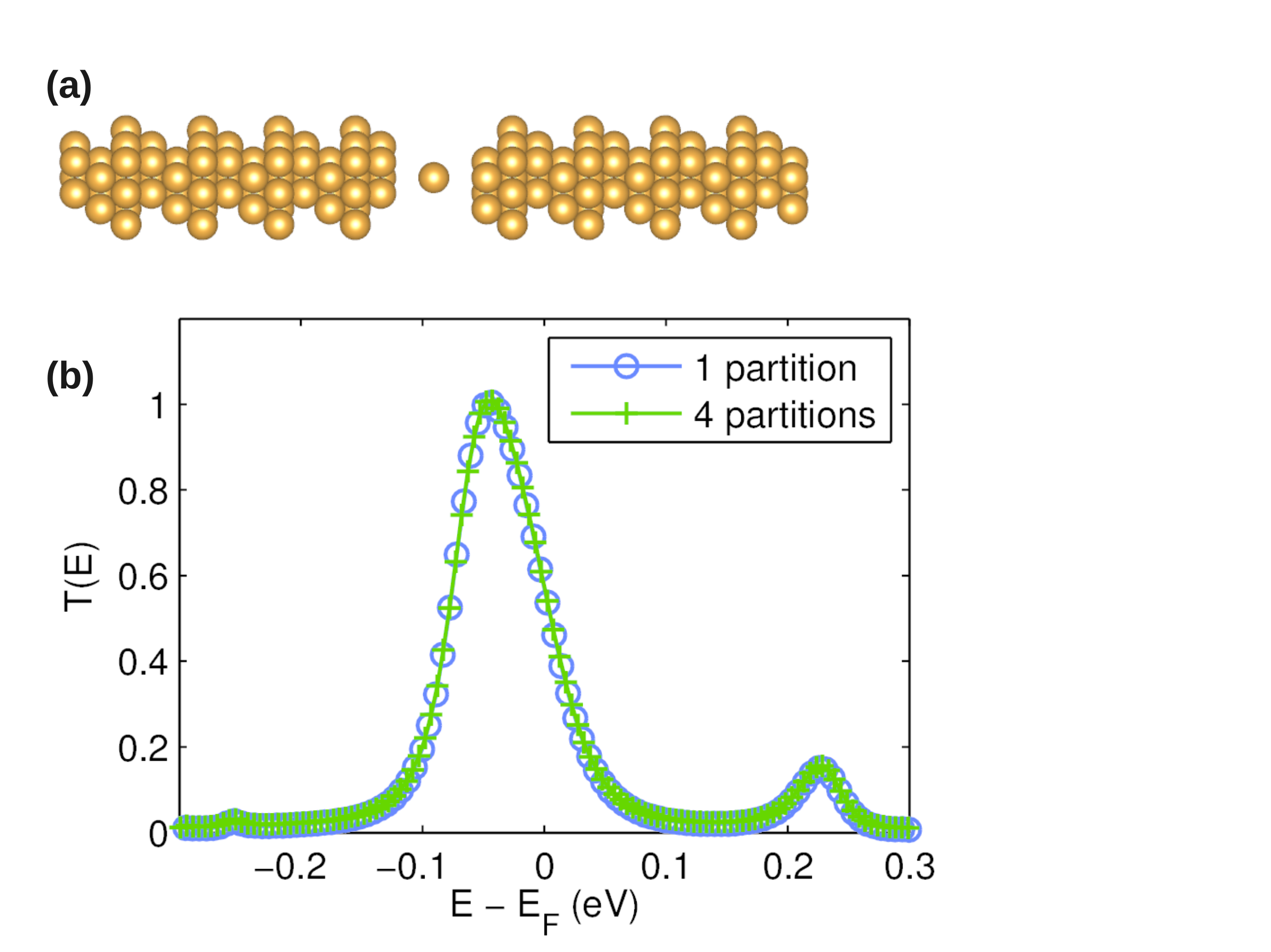} \vspace*{-10mm} 
\end{center}
\caption{
(a) Structure of the Au(111) nanowire/atom/nanowire system considered here and in Ref.~\cite{TRANSEC}.  (b) TRANSEC calculated results for transmission, $T(E)$, for the Au(111) nanowire/atom/nanowire system shown on top.  Results obtained using the partitioned shift-without-invert scheme agree with 
our results obtained with non-partitioned PARPACK in Ref.~\cite{TRANSEC} to better than two decimal places.
\label{fig:Au-results} } 
\end{figure}

To gauge the shift-without-invert method's usefulness in practice, we next applied Algorithm \ref{algo:partition} to one of our primary test systems of Ref.~\cite{TRANSEC}, 
consisting of Au(111) nanowire electrodes with an Au atomic point contact as the scattering region, and a gap of 9.3 $a_0$ between the central Au atom and each lead.  The system's structure, shown in Fig.~\ref{fig:Au-results}(a), is identical to that used in Ref.~\cite{TRANSEC}.  Same as in  Ref.~\cite{TRANSEC}, 
we used Gaussian absorbing potentials centered at the ends of the two electrodes, with strength 100 mRy and standard deviation 8.5 $a_0$; and 
we used a norm-conserving Troullier-Martins pseudopotential for Au with electronic configuration of $5d^{10}6s^{1}6p^0$ and \textit{s}/\textit{p}/\textit{d} cutoff radii of 2.77/2.60/2.84 $a_0$.  
The real-space grid had $N=$ 234,500 grid points, of which the lowest $n_r =$ 2,930 eigenpairs, roughly 1\% of the total, were computed.  The eigenvalues ranged from $E_F \: -$ 0.549 Ry to $E_F \: +$ 1.638 Ry.  As reported in Ref. \cite{TRANSEC}, this single-partition calculation took about $T_{wall,1} =$ 41 wall-clock hours on two nodes of Intel E5-2630 machines, each node with two hex-core CPUs (a total $N_1 =$ 24 cores).  Thus, the total CPU time was $T_{CPU,1} =$ 980 core-hours.  

In the current work, we have performed the same calculation in four separate partitions of six cores (one hex-core CPU) per partition on the same type of processors, giving a total again of $N_{cores} =$ 24 cores.  The total CPU time for the four runs reduced to only $T_{CPU} =$ 320 core-hours, about three times
faster than the single-partition run parallelized only via PARPACK over grid points.  The longest of the four partitioned runs took $\max_j\{ T_{wall,j}\} =$ 17 hours, a factor $\sim$2.5 less elapsed wall-clock time than the PARPACK-only run.   
The partition parameters were $n_{r,j} = (1019,\: 891,\: 675,\: 524)$ and $E_{S,j} - E_F = (-0.045,\: 0.795,\: 1.281,\: 1.535)$ Ry.  
As shown in Fig.~\ref{fig:Au-results}(b), the $T(E)$ results obtained by both methods agree to within 0.5\% of the peak height, 
which is well within the typical margin of error of TRANSEC $T(E)$ calculations.  

To determine the portion of this observed speedup attributable to the partitioning scheme, 
we also performed a new \emph{single}-partition reference calculation for the same system and same $n_r$ with just $N_1 =$ 6 cores.  In this case, the calculation took $T_{wall,1} =$ 52 hours of elapsed wall-clock time, or a total of $T_{CPU,1} =$ 310 core-hours.  Compared to this 6-core reference run, the shift-without-invert scheme with 24 cores exhibited a parallel speedup of $\eta_{_{CPU}} =$ 98\% as measured by CPU time, and $\eta_{wall} =$ 79\% measured by elapsed wall-clock time.  By contrast, the original single-partition TRANSEC algorithm exhibited speedup of just 31\%  for 24 cores compared to the 6-core reference run.  

\subsection{Benzene dithiol (BDT) molecule with Au(111) nanowire electrodes \label{sec:BDT}}

\begin{figure}[!h]
\begin{center}
\includegraphics[scale=0.5]{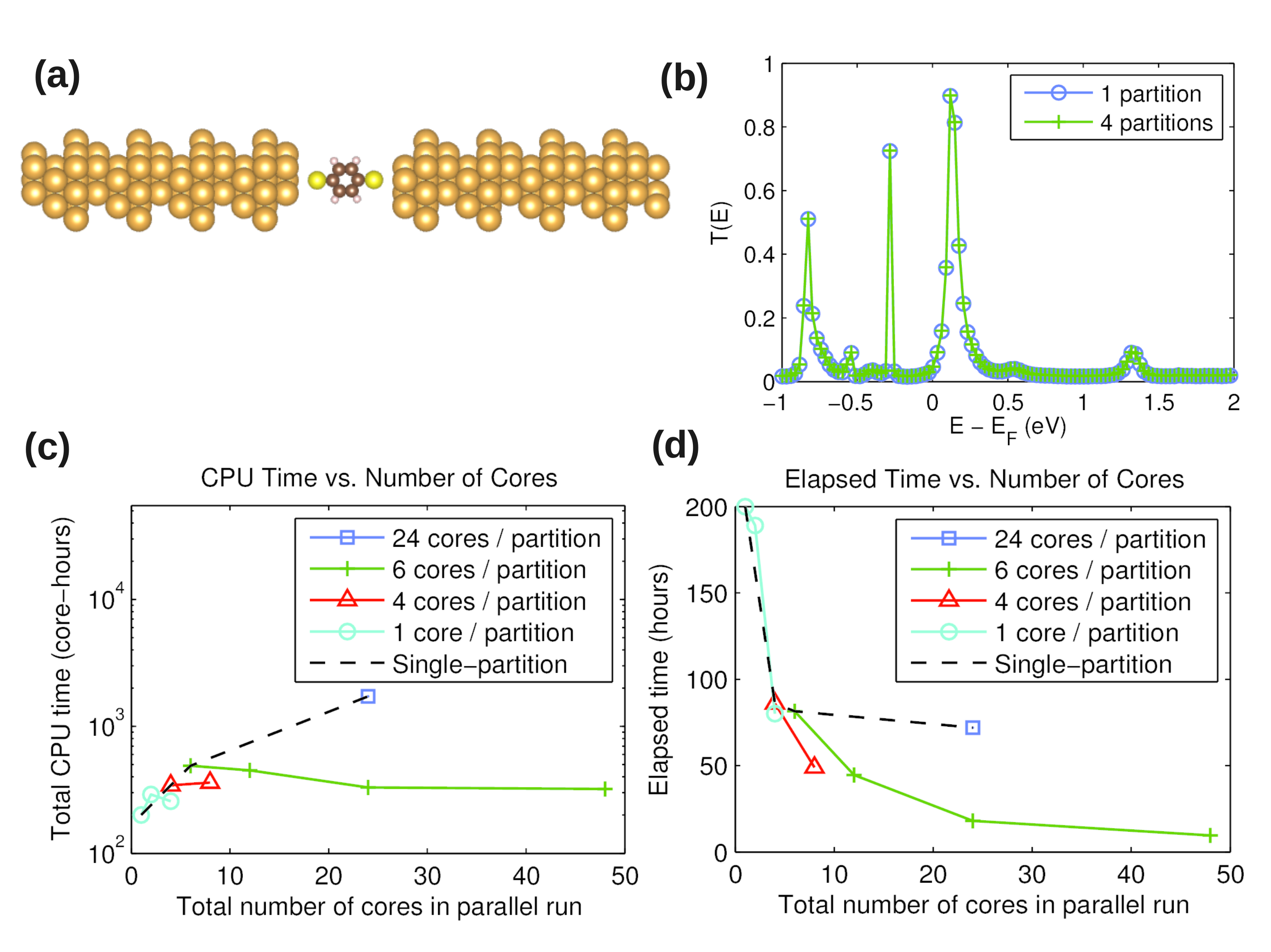}
\end{center}
\caption{(a) Structure of the the benzene dithiol (BDT) molecular junction between Au(111) nanowire electrodes, as in Ref.~\cite{TRANSEC} except with a 6.6 $a_0$ molecule-electrode gap.  (b) Computed $T(E)$ with 4 partitions compared to the standard single-partition TRANSEC.  (c) Total CPU time $T_{CPU}$ vs. number of cores $N_{cores}$ for the $T(E)$ calculation in BDT, shown in semi-log scale.  $T_{CPU}$ shown decreasing with $N_{cores}$ implies super-linear parallel speedup, as discussed in the main text.  (d) Net elapsed time $\max_j\{ T_{wall,j} \}$ (wall-time of longest-running partition) vs. $N_{cores}$. 
Results are shown for runs with 24 (blue square \cite{TRANSEC}), 6 (green pluses), 4 (red triangles), and 1 (cyan circles) cores per partition.  Note the leftmost point on each curve is a standard single-partition run, as in Ref.~\cite{TRANSEC}; the remaining points are multiple-partition runs using the shift-without-invert scheme. \label{fig:BDT-results}
To compare parallelization schemes, one can compare the elapsed time (part (c) of figure) for different runs at a fixed total number of cores in the calculation.  As can be seen from (c), four partitions of six cores each (green pluses) parallelize far better than a single partition of 24 cores (blue square) \cite{TRANSEC}.  }
\end{figure}

We also applied the partitioned shift-without-invert scheme to another of our principal test systems from Ref.~\cite{TRANSEC}, a molecular junction with the same Au(111) nanowire electrodes and absorbing potentials as in Section \ref{sec:au111} above and a benzene dithiol (BDT) molecule as the scattering region.  
The system structure used is identical to that in Ref.~\cite{TRANSEC}, except that there an electrode-molecule gap of 3.2 $a_0$ was used to match a similar calculation by Stokbro \textit{et al.} \cite{Stokbro-BDT}, and here we use a larger 6.6 $a_0$ gap, as shown in Fig.~\ref{fig:BDT-results}(a), to demonstrate the gap-dependence of $T(E)$.  
As in Ref.~\cite{TRANSEC}, we used norm-conserving Troullier-Martins pseudopotentials
with \textit{s}/\textit{p}/\textit{d} cutoff radii of 1.69/1.69/1.69 $a_0$ for S, \textit{s}/\textit{p} cutoff radii of 1.46/1.46 $a_0$ for C, and \textit{s} cutoff radius of 1.28 $a_0$ for H.
The real-space grid had $N=$ 257,000 grid points, of which the lowest $n_r =$ 3,210 eigenpairs, about 1\% of the total, were computed.  These eigenvalues ranged from $E_F \: -$ 0.977 Ry to $E_F \: +$ 1.670 Ry.  The single-partition calculation took about $T_{wall,1} =$ 72 hours on $N_1 =$ 24 cores (two nodes) of Intel E5-2630, for a total of $T_{CPU,1} =$ 1,730 CPU core-hours.  

\begin{table}[htbp]
	\centering
		\begin{tabular}{|c|c|c|c|c|c|} \hline
			$p$ &	$N_{cores} / p$ &	$\{ n_{r,j} \}$ & $\{ E_{S,j} \} - E_F$ (Ry) &	$\{ T_{wall,j} \}$ (hours) &	$T_{CPU}$ (core-hours) \\	\hline
			1 & 1 & 3209 & 0.34 & 200 & 200 \\
			1 & 4 & 3209 & 0.34 & 86 & 344 \\
			1 & 6 & 3209 & 0.34 & 82 & 489 \\
			1 & 24 & 3209 & 0.34 & 72 & 1728 \\ \hline
			2 & 1 & $(2252, 1475)$ & $(0.06, 1.40)$ & $(100, 193)$ & 293 \\
			2 & 4 & $(2252, 1475)$ & $(0.06, 1.40)$ & $(41, 50)$ & 362 \\ 
			2 & 6 & $(2252, 1475)$ & $(0.06, 1.40)$ & $(30, 43)$ & 438 \\ \hline
			4 & 1 & $(1188, 1035, 807, 655)$ & $(-0.24, 0.82, 1.29, 1.56)$ & $(23, 77, 79, 77)$ & 256 \\ 
			4 & 6 & $(1188, 1035, 807, 655)$ & $(-0.24, 0.82, 1.29, 1.56)$ & $(6, 18, 16, 15)$ & 330 \\ \hline
			8 & 6 & $(561, 520, 485, 464, $ & $(-0.58, 0.14, 0.65, 0.96,$ & $(2, 4, 5, 6, 8, 9, 10, 9)$ & 321 \\
			  &   &  446, 466, 440, 376) & 1.18, 1.37, 1.53, 1.64) & & \\
			\hline
		\end{tabular}
	\caption{Details of partitioned and non-partitioned BDT calculations, including number of partitions $p$, number of cores per partition, partition parameters, and timing information.}
	\label{tab:BDT}
\end{table}

As in Section~\ref{sec:au111}, we performed the same $T(E)$ calculation with the shift-without-invert scheme, using four partitions on six cores (one hex-core CPU) per partition, 
with the same type of nodes, for a total $N_{CPU} =$ 24 cores.  
As shown in Fig.~\ref{fig:BDT-results}(b), the partitioned and non-partitioned $T(E)$ agree to within the margin of error of TRANSEC calculations.  
The total CPU time for the four partitioned runs  
reduced to just $T_{CPU} =$ 330 CPU core-hours, a factor $>$5 less than the single-partition run using PARPACK 
on 24 cores.  
The longest of the four partitioned runs took $\max_j\{ T_{wall,j} \} =$ 18 wall-clock hours, 
a factor $\sim$4 savings in elapsed time compared to the non-partitioned PARPACK run on the same number of cores.  
In addition, the overhead of the 
partitioned runs, i.e., combining the results together by removing redundancy, 
took $T_{CPU} =$ 50 core-hours.  

To further investigate the method's parallel performance in large-scale calculations, we performed a series of $T(E)$ calculations with one, four, and six cores per partition.  
Details of these BDT calculations, including the partition parameters, are given in Table \ref{tab:BDT}.  
We have summarized the speedup data, our primary result in this work, in Fig.~\ref{fig:BDT-results}(c), showing $T_{CPU}$ vs. $N_{cores}$ and Fig.~\ref{fig:BDT-results}(d), showing $\max_j\{T_{wall,j} \}$ vs. $N_{cores}$.  
Each curve shown has a fixed number of cores per partition, $N_{cores}/p$. 
Thus 
one can compare parallelization schemes by evaluating the timings for various curves 
at a fixed $N_{cores}$ position along the horizontal axis.  
The number $p$ of partitions is of course given by $N_{cores}$ divided by $N_{cores}/p$.  

The left-most data point displayed in each curve is always a single-partition calculation (serial or parallelized over grid points).  
Thus, one can visualize the parallel efficiency of the shift-without-invert scheme by comparing the $T_{CPU}$ trend of each curve to the $T_{CPU}$ value of the left-most data point.  
The runs with six cores per partition (shown as green pluses) exhibit $T_{CPU}$ decreasing with $N_{cores}$, or equivalently parallel speedup $\eta_{_{CPU}} >$ 100\%.  For all the curves, $T_{CPU}$ is constant or at most weakly increasing, or equivalently $\eta$ is near 100\% or even better.  
For example, comparing the four-partition \textit{vs.} single-partition runs with $N_{cores}/p=$ 6 cores per partition, we see that parallel speedup was $\eta_{_{CPU}}=$ 150\% by CPU time and $\eta_{wall}=$ 120\% by elapsed time.  
Moreover, the single-partition run with 24 cores (shown as a blue square) from Ref.~\cite{TRANSEC} has $T_{CPU}$ and $T_{wall}$ far greater than the shift-without-invert results with the same $N_{cores}$.  
Although the original TRANSEC method has been designed to handle very large calculations, in this case our original single-partition result with 24 cores was clearly over-parallelized for the size of the given calculation.  

The dashed line connects the left-most point of each curve, and so represents the speedup trend of the standard TRANSEC algorithm with parallelization only over grid points \cite{TRANSEC}.  
Moreover, contrasting the trend of the dashed curve to each solid curve vividly 
illustrates how the limitations of PARPACK-only parallelization can be overcome by shift-without-invert partitioning of the eigenspace.  
These speedup results are particularly noteworthy because the BDT junction between Au(111) nanowire electrodes is 
a challenging nano quantum transport system.

\section{Conclusions}

We have developed a partitioned shift-without-invert scheme that significantly improves the performance of large iterative partial diagonalization algorithms.  This scheme can theoretically reduce the computational cost from $O(N n_r^2)$ to $O(N n_r)$, 
as noted in Eq.~\ref{eq:linear-complexity}.  
In practice,   
we have illustrated with $n_r/N =$ 10\% (Section~\ref{sec:C-chain}) that the shift-without-invert time $T_{CPU}$ can indeed be less than the single-core time $T_0$, and even with $n_r/N \approx$ 1\% (green pluses in Fig.~\ref{fig:BDT-results}(c)) less than the single-partition time $T_{CPU,1}$.  
The proposed scheme adds another level of parallelization (over the spectrum and the spatial grid), which
provides significant improvement over the already good 
parallelization (over spatial grid only) implemented in TRANSEC. 
As a result, even with the non-optimized partitions and parameters, 
we have readily obtained a factor $>$5 improvement in CPU time for a large TRANSEC calculation by switching from one to four partitions with the same number of cores, as shown in Fig.~\ref{fig:BDT-results}(c).   
The shift-without-invert scheme provides a far simpler overall structure of partitioned methods. 
This is particularly true when compared with partitioned approaches that utilize inversion, 
which require significant additional effort related 
to solving shifted linear equations.  
Moreover, our scheme is applicable to non-Hermitian problems, for such eigenvalue problems very few parallel
algorithms have been proposed.
Finally, our scheme also enables continuation runs, so that previously converged eigenpairs need not
be discarded, but instead we simply place new shifts in unexplored regions of the spectrum to compute desired new eigenpairs. 
The shift-without-invert scheme is expected to be applicable to a wide range of iterative 
eigenvalue problems: since we base our partitioned solver on PARPACK, it inherits the
remarkable robustness and generality of the PARPACK package.  
Thus for a wide range of eigenvalue problems where ARPACK/PARPACK is applicable, one can adapt our 
partitioned scheme to improve parallel scalability.

\section{Acknowledgements}

We would like to thank Leeor Kronik and Oded Hod for helpful comments and discussions.  This research was partially supported by the European Research Council, the Israel Science Foundation, and the NSF via grants DMS-1228271 and DMS-1522587.

\bibliographystyle{unsrt} 
\bibliography{shift_no_invert0}

\begin{thebibliography}{10}

\bibitem{TRANSEC}
B.~Feldman, T.~Seideman, O.~Hod, and L.~Kronik.
\newblock Real-space method for highly parallelizable electronic transport
  calculations.
\newblock {\em Phys. Rev. B}, 90:035445, 2014.

\bibitem{cts:94}
J.~R. Chelikowsky, N.~Troullier, and Y.~Saad.
\newblock Finite-difference-pseudopotential method: Electronic structure
  calculations without a basis.
\newblock {\em Phys. Rev. Lett.}, 72:1240--1243, 1994.

\bibitem{parsec-review}
L.~Kronik, A.~Makmal, M.~L. Tiago, M.~M.~G. Alemany, M.~Jain, X.~Huang,
  Y.~Saad, and J.~R. Chelikowsky.
\newblock {\em Phys.~Status Solidi B}, 243:1063, 2006.

\bibitem{Meerbergen:1997:IRA}
K.~Meerbergen and A.~Spence.
\newblock Implicitly restarted {Arnoldi} with purification for the shift-invert
  transformation.
\newblock {\em Math. Comp.}, 66(218):667--689, 1997.

\bibitem{appinv-eig00}
L.~Bergamaschi, G.~Pini, and F.~Sartoretto.
\newblock Approximate inverse preconditioning in the parallel solution of
  sparse eigenproblems.
\newblock {\em Numer. Linear Algebra Appl.}, 7:99 -- 116, 2000.

\bibitem{sip07}
H.~Zhang, B.~Smith, M.~Sternberg, and P.~Zapol.
\newblock {SIPs: Shift-and-Invert Parallel Spectral Transformations}.
\newblock {\em ACM Trans. Math. Software}, 33(2):article 9, 2007.

\bibitem{silan-14}
H.M. Aktulga, L.~Lin, C.~Haine, E.G. Ng, and C.~Yang.
\newblock Parallel eigenvalue calculation based on multiple shift-invert
  {Lanczos} and contour integral based spectral projection method.
\newblock {\em Parallel Computing}, 40(7):195--212, 2014.

\bibitem{Santra}
R.~Santra and L.~S. Cederbaum.
\newblock Non-hermitian electronic theory and applications to clusters.
\newblock {\em Phys. Rep.}, 368(1):1--117, 2002.

\bibitem{CSYM}
A.~Bunse-Gerstner and R.~St{\"o}ver.
\newblock On a conjugate gradient-type method for solving complex symmetric
  linear systems.
\newblock {\em Lin. Alg. Appl.}, 287:105--123, 1999.

\bibitem{complex-Lanczos}
R.~W. Freund.
\newblock Conjugate gradient-type methods for linear systems with complex
  symmetric coefficient matrices.
\newblock {\em SIAM J. Sci. Stat. Comput.}, 13:425--448, 1992.

\bibitem{SaadBook}
Y.~Saad.
\newblock {\em Iterative Methods for Sparse Linear Systems}.
\newblock SIAM Press, 2003.

\bibitem{benzi-precond02}
M.~Benzi.
\newblock Preconditioning techniques for large linear systems: {A} survey.
\newblock {\em J. Comput. Phys.}, 182(2):418--477, 2002.

\bibitem{arpack}
R.~B. Lehoucq, D.~C. Sorensen, and C.~Yang.
\newblock {\em {ARPACK} User's Guide: {S}olution of Large Scale Eigenvalue
  Problems with Implicitly Restarted {Arnoldi} Methods}.
\newblock SIAM, 1998.

\bibitem{govl:96}
G.~H. Golub and C.~F.~Van Loan.
\newblock {\em Matrix Computations}.
\newblock Johns Hopkins University Press, Baltimore, MD, 3rd edition, 1996.

\bibitem{pchefsi}
Y.~Zhou, Y.~Saad, M.~L. Tiago, and J.~R. Chelikowsky.
\newblock Parallel self-consistent-field calculations using
  {C}hebyshev-filtered subspace acceleration.
\newblock 74(6):066704, 2006.

\bibitem{slicing-12}
G.~Schofield, J.R. Chelikowsky, and Y.~Saad.
\newblock A spectrum slicing method for the {Kohn-Sham} problem.
\newblock {\em Comp. Phys. Comm.}, 183:497--505, 2012.

\bibitem{lanczos-para87}
S.~W. Bostic and R.~E. Fulton.
\newblock {Implementation of the Lanczos method for structural vibration
  analysis on a parallel computer}.
\newblock {\em Computers \& Structures}, 25(3):395--403, 1987.

\bibitem{davis-book}
T.~Davis.
\newblock {\em Direct Methods for Sparse Linear Systems}.
\newblock Fundamentals of Algorithms. SIAM Press, Philadelphia, 2006.

\bibitem{mumps}
P.~R. Amestoy, I.~S. Duff, J.-Y. L'Excellent, and J.~Koster.
\newblock A fully asynchronous multifrontal solver using distributed dynamic
  scheduling.
\newblock {\em SIAM J. Matrix Anal. Appl.}, 23(1):15--41, 2001.

\bibitem{temp:lin}
R.~Barrett, M.~W. Berry, T.~F. Chan, J.~Demmel, J.~Donato, J.~Dongarra,
  V.~Eijkhout, R.~Pozo, C.~Romine, and H.~van~der Vorst, editors.
\newblock {\em Templates for the solution of linear systems}.
\newblock SIAM, Philadelphia, PA, 1994.

\bibitem{petsc-ug}
S.~Balay, S.~Abhyankar, M.~F. Adams, J.~Brown, P.~Brune, K.~Buschelman,
  L.~Dalcin, V.~Eijkhout, W.~D. Gropp, D.~Kaushik, M.~G. Knepley, L.~Curfman
  McInnes, K.~Rupp, B.~F. Smith, S.~Zampini, and H.~Zhang.
\newblock {PETS}c users manual.
\newblock Technical Report ANL-95/11 - Revision 3.6, Argonne National
  Laboratory, 2015.

\bibitem{website:arpack-ng}
The {ARPACK-NG} project.
\newblock \url{http://github.com/opencollab/arpack-ng}.

\bibitem{sorens:92}
D.~C. Sorensen.
\newblock Implicit application of polynomial filters in a $k$-step {Arnoldi}
  method.
\newblock {\em SIAM J. Matrix Anal. Appl.}, 13:357--385, 1992.

\bibitem{spectral-dos}
L.~Lin, Y.~Saad, and C.~Yang.
\newblock Approximating spectral densities of large matrices.
\newblock {\em SIAM Review}, 58:34--65, 2016.

\bibitem{int-count}
E.~Di~Napoli, E.~Polizzi, and Y.~Saad.
\newblock Efficient estimation of eigenvalue counts in an interval.
\newblock {\em ArXiv e-prints, arXiv:1308.4275}, 2013.

\bibitem{Stokbro-BDT}
K.~Stokbro, J.~Taylor, M.~Brandbyge, J.-L. Mozos, and P.~Ordej\'{o}n.
\newblock {Theoretical study of the nonlinear conductance of Di-thiol benzene
  coupled to Au(111) surfaces via thiol and thiolate bonds}.
\newblock {\em Comp. Mat. Sci.}, 27:151, 2003.

\end{thebibliography}
\end{document}